\documentclass[aps,prl,twocolumn,showpacs,groupedaddress]{revtex4}
\usepackage{epsfig}
\usepackage{bm}
\newcommand{\ds}{\displaystyle}

\begin{document}
\preprint{02-02}
\title{Effect of gluon-exchange pair-currents on the ratio
	$\mu_p G_E^p/G_M^p$}

\author{Murat M. Kaskulov}
 \email{kaskulov@pit.physik.uni-tuebingen.de}
\author{Peter Grabmayr}
 \email{grabmayr@uni-tuebingen.de}
\affiliation{Physikalisches Institut, Universit\"at  T\"ubingen,
		 D-72076 T\"ubingen, Germany}
\date{\today}

\begin{abstract}
  The effect of one-gluon-exchange (OGE) pair-currents on the ratio~$\mu_p
  G_E^p/G_M^p$ for the proton is investigated within a nonrelativistic
  constituent quark model (CQM) starting from $SU(6) \times O(3)$ nucleon wave
  functions, but with relativistic corrections.  We found that the OGE
  pair-currents are important to reproduce well the ratio~$\mu_p G_E^p/G_M^p$.
  With the assumption that the OGE pair-currents are the driving mechanism for
  the violation of the scaling law we give a prediction for the ratio~$\mu_n
  G_E^n/G_M^n$ of the neutron.
\end{abstract}

\maketitle

{\it Introduction:}~~~ 
Recently the ratio $\mu_p G_E^p/G_M^p$  between the electric $G^p_E(Q^2)$ and
magnetic $G^p_M(Q^2)$ form factors of the proton has been extracted from
experimental data on the recoil proton polarization in elastic electron
scattering with polarized electrons up to
$Q^2\sim$5~GeV$^2$~\cite{Milbrath1999,Jones2000,Gayou2001,Gayou2002}.  These
experiments are of importance because they are direct measurements of the form
factor ratio, and the present results are in contradiction to previous
analyses~\cite{Milbrath1999,Jones2000,Gayou2001,Gayou2002,Brash2002}.
Historically, the determination of the electric and magnetic form factors up
to several GeV were based on the Rosenbluth separation, and they were found
compatible with the scaling laws:
\begin{equation}\label{scaling} 
	G_E^p(Q^2) = G_M^p(Q^2)/{\mu_p} =  G_{D}(Q^2)\ \ . 
\end{equation} 
where $G_{D}(Q^2)$ represents the dipole form factor. 
 
The form factors and particularly the ratio give insight to the main features
of the dynamical processes and are very useful for a test of the nucleon
models~\cite{Thomas_book}.  The remarkable feature of the new experimental
data is that they show a decrease of the ratio~$\mu_{p} G_E^p/G_M^p$ from
unity, indicating a significant deviation from this simple scaling law, but
also from the simple constituent quark model.

Within different hadronic models the calculations for the proton ratio $\mu_p
G_E^p/G_M^p$ became available, with Ref.~\cite{Frank2} presenting one of the
earliest. We will restrict this discussion to the most recent calculations
which agree reasonably well with the trend of the experimental data and which
will allow to make predictions at higher $Q^2$ than the present data.  In the
cloudy bag model (CBM)~\cite{Thomas}, the pion field required by chiral
symmetry is quantized and coupled to the MIT bag~\cite{MIT1}.  Addition of the
pion cloud improves the MIT bag model results~\cite{Lu1}, in which the
decrease of $\mu_p G_E^p/G_M^p$ is an inherent property.  It was shown for a
CBM formulated on the light cone~\cite{Miller}, that the combination of
Poincar\'e invariance and pion effects is sufficient to describe $\mu_p
G_E^p/G_M^p$.  Several groups have studied different effects within CQMs.  In
the Goldstone boson exchange CQM~\cite{GlozmanRiska} the baryon is considered
as a system of three constituent quarks with an effective $qq$ hyperfine
interaction mediated by the octet of pseudoscalar mesons.  This model together
with the point-form spectator approximation~\cite{GlozmanBoffi}, which
provides a covariant framework, leads to a rather close description of the
nucleon form factors and the available $\mu_p G_E^p/G_M^p$ data.  Calculations
of Ref.~\cite{Cardarelli} performed within CQM and light-front formalism,
showed that a suppression of the ratio can be expected in the CQM, if the
relativistic effects generated by kinematical $SU(6)$ breaking due to the
Melosh rotation of the constituent spins are taken into account.  Finally, the
most recent calculations based on relativistic quark models are from
Ref.~\cite{MillerFrank}, where the hadron helicity nonconservation induced by
the Melosh transformation was recognised to affect the ratio.  The
implementation of relativity is an common feature of all these works and all
emphasize the necessity of both kinematical and dynamical relativistic
corrections for the interpretations of the decrease of the ratio~$\mu_p
G_E^p/G_M^p$.

In the non-relativistic constituent quark model~(NRCQM)~\cite{Isgur}, the
effective degrees of freedom are the massive quarks moving in a
self-consistent potential whose specific form is dictated by considerations of
QCD. Other degrees of freedom like Goldstone bosons or gluons are not
considered in the original version and effectively absorbed into the
constituent quarks.

Theoretically, the explicit introduction of the additional degrees of freedom
in the nucleon structure will change its properties compared to expectations
based on simple quark models in which the baryon is described as a three-quark
state only.  Among different improvements to the naive CQM which could be
essential for dynamical properties of the nucleons, the most important ones
are relativistic kinematical corrections, the introduction of a mesonic cloud
via pion-loop corrections, and dynamical corrections due to the interaction
currents and to the creation of quark-antiquark ($q \bar q$) pairs.  For low
momentum transfer, $q \bar q$ pairs (sea-quarks) are dominant and the mesonic
degrees of freedom become increasingly important.  However, in a recent
study~\cite{Geiger} on ``un-quenching'' the quark model, strong cancellations
between the hadronic components of the $q \bar q$ sea were found which tend to
make the nucleon transparent to photons. These studies provide a natural way of
understanding the success of the valence quark model even though the $q\bar q$
sea is very strong.  At higher momentum transfer and in the presence of residual
$qq$ interaction, the e.m. operators must be supplemented by the
two-body exchange currents.  The inclusion of two-body terms leads beyond the
single-quark impulse approximation, and in dependence on the model for the
$qq$ interaction effectively represents the gluonic or mesonic exchange
degrees of freedom in the e.m. current operator.  In this sense the physical
picture should be similar to nuclear physics, where at low momentum transfer
the nucleons are reasonable degrees of freedom, but at higher momentum
transfer the meson-exchange currents play a prominent 
role~\cite{Kaskulov:2002mc}.

In this work we continue our studies~\cite{Grabmayr1} of the possible role of
interaction currents, in particular OGE pair-currents, for the e.m. properties
of the nucleon.  We use the NRCQM with relativistic corrections, coming from
the Lorentz boost of the nucleon wave function, together with gluonic
corrections for the calculation of the proton ratio~$\mu_{p} G_E^p/G_M^p$ at
momentum transfers beyond 1~GeV$^2$, where effects of the soft pionic cloud
 should be less important.  We show that gluonic corrections to the CQM
are important, and that the ratio~$\mu_{p} G_E^p/G_M^p$ is well reproduced by
the $SU(6) \times O(3)$ wave function of the nonrelativistic quark model.  

{\it The nucleon in the NRCQM:}~~~ 
In the quark model, baryons are considered as three-quark configurations. The
ground state has positive parity with all three quarks in their lowest state,
and the total angular momentum (isospin) of baryons is obtained by
appropriately combining the quark spins (isospins).  In the NRCQM~\cite{Isgur}
a baryon is treated as a non-relativistic three-quark system, and in the
simplest case of equal quark masses~$m_q$ it is described by the Hamiltonian:
\begin{eqnarray}
\label{H3q}
\mathcal{H}_{3q} &=&~~\ds\sum_{i=1}^{3}%
			\Big(m_q + \frac{\ds {\bf p}_i^2}{\ds 2 m_q} \Big) -  \
			\frac{\ds {\bf P}^2}{\ds 6 m_q} \nonumber \\
  && +   	\ds\sum_{i<j}^{3} V^{(conf)}({\bf r}_i,{\bf r}_j)
		  + \ds \sum_{i<j}^{3} V^{(res)}({\bf r}_i,{\bf r}_j) 
\end{eqnarray}
where ${\bf r}_i$, ${\bf p}_i$ are the spatial and momentum coordinates of the
$i$-th quark, respectively, and {\bf P} is the centre-of-mass momentum. The
Hamiltonian ${\cal H}_{3q}$ consists of the nonrelativistic kinetic energy, a
confinement potential $V^{(conf)}$, and a residual interaction
$V^{(res)}$. Here, we take a two-body harmonic oscillator (h.o.) confinement
potential:
$
V^{(conf)}({\bf r}_i,{\bf r}_j) ~\sim~ {\bf \lambda}_i \cdot {\bf \lambda}_j
( {\bf r}_i - {\bf r}_j)^2,~
$
where ${\bf \lambda}_i$ are the Gell-Mann colour matrices of the $i$-th quark,
with $\left<{\bf \lambda}_i \cdot {\bf \lambda}_j\right> = -8/3$ for a $qq$
pair in a baryon.

The phenomenological residual interaction $V^{(res)}$ can be based on various
$qq$ potentials~\cite{GlozmanRiska,Isgur}, which reflect the symmetries and
properties of QCD.  Up to now, its dynamical origin is rather uncertain.  We
use a standard OGE interaction, the strength of which is
determined by the strong coupling constant~$\alpha_s$.  However, unlike
perturbative QCD, where the strong coupling constant~$\alpha_s$ goes to zero
at large inter-quark momenta, we take $\alpha_s$ of the NRCQM as an effective
momentum independent constant.

We start from the simplest form of the NRCQM, i.e. without configuration
mixing, in which the nucleon $| N \rangle$ is described by the lowest
h.o. three quark configurations $(0s)^3[3]_X$ in the translationally-invariant
shell model (TISM):
\begin{equation}
| N \rangle = \Big|(0s)^3 [3]_X L=0, ST =
\frac{1}{2} \frac{1}{2} [3]_{ST},~ J^P = \frac{1}{2}^{+} \Big\rangle 
\end{equation}
where the colour part is omitted.  After having removed the centre-of-mass
coordinate ${\bf R}$ from the TISM configuration, the ground state
eigenfunction depends only on the Jacobi relative coordinates ${\bf \rho}_1$
and ${\bf \rho}_2$ of the quarks:
\begin{equation}
| (0s)^3 ({\bf \rho}_1,{\bf \rho}_2) \rangle 
\sim
\exp \left( - \frac{1}{4 b^2} {\bf \rho}^2_1 - \frac{1}{3 b^2} {\bf \rho}^2_2
\right)
\end{equation}
where the constant~$b$~determines the average hadronic size of the baryon.
Note that the elimination of ${\bf R}$ is crucial for correctly counting the
baryonic states. This is one reason why the nonrelativistic approach is so
successful in spectroscopy.

{\it The nucleon e.m. Sachs form factors:}~~~
The nucleon e.m. form factors are functions of the square of the momentum
transfer in the scattering process ${Q}^2 = -q^{\mu} q_{\mu}$.  The Sachs form
factors, $G_{E(M)}$, fully characterize the charge and current distributions
inside the nucleon~\cite{Sachs1} and can be written in terms of Dirac and
Pauli form factors~$\mathcal{F}_1$ and~$\mathcal{F}_2$, respectively.  The
most general form of the nucleon e.m. operator $J^{\mu}_{em}(x)$, which
defines $\mathcal{F}_1$ and $\mathcal{F}_2$, satisfies the requirements of
relativistic covariance and the condition of gauge invariance; it is of the
form
\begin{eqnarray}
\langle N(p',s') | J^{\mu}_{em}(0) | N(p,s) \rangle &=&   \\ \nonumber
 \bar{u}({\bf p}',s') \Big[ \gamma^{\mu} \mathcal{F}_{1}(Q^2) &+&
i\frac{\sigma^{\mu\nu}q_{\nu}}{2 M_N}\mathcal{F}_{2}(Q^2) \Big] u({\bf p}, s) ,
\end{eqnarray}
with $q^{\nu} = p'^{\nu} - p^{\nu}$.  The Breit frame, where the incoming
momentum ${\bf p} = - {\bf q}/2$ is scattered to the momentum ${\bf p}' = {\bf
q}/2$, is characterized by ${Q}^2 = {\bf q}^2$. In this frame the nucleon
electric $G_E$ and magnetic $G_M$ form factors can be interpreted as Fourier
transforms of the distributions of charge and magnetization, respectively:
\begin{eqnarray}
\Big< N_{s'}(\frac{{\bf q}}{2})\Big|~{\bf J}_{em}(0)~
  \Big| N_s(-\frac{{\bf q}}{2}) \Big> &=&%
\chi^{\dagger}_{s'}\frac{i{\bf\sigma}\times{\bf q}}{2 M_N}
\chi_s G_M(q^2) ~~~~\\
\Big< N_{s'}(\frac{{\bf q}}{2}) \Big| ~J^{0}_{em}(0)~
  \Big| N_s(-\frac{{\bf q}}{2}) \Big> &=&%
   \chi^{\dagger}_{s'} \chi_s G_E(q^2) \ \
\end{eqnarray}
where $\chi^{\dagger}_{s'}$ and $\chi_s$ are Pauli spinors for the initial and
final nucleons.

Starting from the rest frame, the spherical nucleon is expected to undergo a
Lorentz contraction along the direction of motion.  Results of previous
studies suggest that the consistent treatment of the form factors should be
supplemented by the relativistic boost~\cite{Wagenbrunn}.  But a complete
solution of a covariant many-body problem is difficult; the use of the
light-cone dynamics~\cite{Chung} for constituent quarks leads to the
introduction of additional parameters. Thus, a semiclassical prescription
proposed in Ref.~\cite{Licht} and successfully applied in a CBM~\cite{Lu1} is
used here. Thereby, the relativistic form factors can be derived in the Breit
frame from the corresponding nonrelativistic ones by a simple substitution:
\begin{equation}
\label{Lboost}
G_{E(M)}(Q^2) \to \eta  G_{E(M)}(\eta Q^2),
\end{equation}
where $ \ds \eta = {M^2_N}/{E^2_N} $ and $ E_N^2 = M_N^2 + {{\bf q}^2}/{4}$.
The scaling factor $\eta$ in the argument of $G_{E(M)}$ arises from the
coordinate transformation of the struck quark, and the pre-factor in
Eq.(\ref{Lboost}) comes from the reduction of the integral measure of the two
spectator quarks in the Breit frame. This simple boost together with the
NRCQM nucleon wave function does not addmix configurations
with nonzero orbital angular momentum; it leads to the hadron helicity
conserving solution.  Note, that imposing Poincar\'e invariance in a
relativistic CQM causes substantial violation of the helicity conservation
rule~\cite{MillerFrank}, and results in an asymptotic behaviour of form
factors which differs from that as expected in pQCD~\cite{Lepage}.

We first consider the nucleon single-quark current ${j}_{q_i}^{\mu}(x)$
contribution:$J^{\mu}_{em}(x) = \sum_{i=1}^{3} {j}_{q_i}^{\mu}(x).$
In the CQM the e.m. vertex of the internal quarks should be assumed to have a
spatially extended structure that may be described by a form
factor~$F_{q}({\bf q}^2)$. The most general form for the covariant e.m. current
operator of the constituent quarks is written as~\cite{Gross}:
\begin{equation}
\label{J_3q_Modif}
{j}_{q_i}^{\mu}(x) = \mathcal{Q}_i \bar{q}_i(x)
\Big\{ \gamma^{\mu} + \Big(F_q({\bf q}^2) - 1 \Big)
\Big[\gamma^{\mu} - \frac{\gamma \cdot q q^{\mu}}{q^2} \Big] \Big\} q_i(x),
\end{equation}
where $q_i(x)$ is the quark field operator, $\mathcal{Q}_i$ is its charge in
units of $e$:~
$\mathcal{Q}_i = 1/2 \left[ 1/3 + \tau^3_i \right].$
This vertex, in which the first term corresponds to pointlike quarks,
maintains the requirement of current conservation, as the form factor
modification appears only in a purely transverse term.  The nonrelativistic
reduction of Eq.(\ref{J_3q_Modif}) for pointlike quarks,
$F_{q}({\bf q}^2)=1$, leads to the standard one-body e.m. current operators:
$\hat{\rho}_{3q}({\bf q})
= \sum_{i=1}^{3} \mathcal{Q}_i e^{i \bf{q} \cdot {\bf r}_i}$
and
$\hat{\bf j}_{3q}({\bf q})
= \frac{1}{2 m_q}
\sum_{i=1}^{3} \mathcal{Q}_i e^{i \bf{q} \cdot {\bf r}_i}
\Bigl( {\bf p}'_i + {\bf p}_i + i {\bf \sigma}_i \times {\bf q} \Bigr), ~
$
where we have retained only the lowest order contributions.  This is in spirit
of a NRCQM, where the main contribution to the e.m. moments is expected to
come from the non-relativistic single quark currents, which by the choice of
the effective quark mass already incorporates substantial relativistic
corrections~\cite{Buchmann}.  It follows that one should not use
next-to-leading order relativistic corrections proportional to $\sim {\bf
q}^2/8 m_q^2 $ in the charge operator $\hat{\rho}_{3q}({\bf q})$,
for example the Darwin-Foldy term, if one ignores them in the kinetic energy.

\begin{figure}[h]
\begin{center}
\epsfig{file=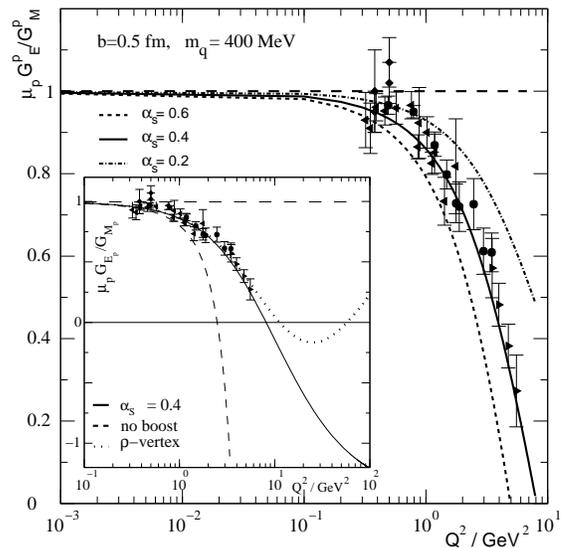,width=0.85\columnwidth,clip} 
\caption{\label{fig:gegmprot4}
The ratio~$\mu_{p} G^p_{E}/G^p_{M}$ for the proton is calculated within the
 NRCQM for $m_q$=400~MeV and $b$=0.5~fm and 3 values of $\alpha_s$.
 The interaction is assumed to be pointlike.
 The calculations are compared to
 data from Refs.~\cite{Jones2000,Milbrath1999,Gayou2001,Gayou2002}.
 The insert shows the ratio over an extended range of $Q^2$ for the best value
 $\alpha_s$=0.4, without Lorentz boost (dashed line) or with a $\rho$ vertex
 form factor (dotted line).
 }
\end{center}
\end{figure}

The naive CQM results in the following nucleon e.m. form factors~$G^{(3q)}_E$
and~$G^{(3q)}_M$:

\begin{eqnarray} 
\label{OB_E}
G^{(3q)}_E ({\bf q}^2) &=& 
e_N \exp \left(-{\bf q}^2  b^2/6 \right)  \\
\label{OB_M}
G^{(3q)}_M ({\bf q}^2) &=&  \frac{M_N}{m_q}
\ \mu_N   \exp \left(-{\bf q}^2 b^2/6\right)  
\end{eqnarray} 
where $e_N$ and $\mu_N$ are the charge and CQM magnetic moment of the nucleon:
$e_N =  \frac{1}{2} \langle N |(1 + \tau_3) | N \rangle, ~
\mu_N = \frac{1}{6} \langle N | ( 1 + 5 \tau_3 ) | N \rangle .$
Due to the same momentum dependence, Eqs.(\ref{OB_E}) and~(\ref{OB_M}) lead
to the scaling law noted in Eq.(\ref{scaling}); a ratio of unity is obtained
as presented by the long dashed line in
Fig.~\ref{fig:gegmprot4}. Clearly, the scaling law is in contradiction with the
recent proton experiments~\cite{Jones2000,Milbrath1999,Gayou2001,Gayou2002}.

{\it The OGE pair-current:}~~
In the presence of residual OGE interactions between the quarks the total
current operator of the hadron cannot simply be a sum of free quark currents,
but must be supplemented by two-body currents. These two-body currents are
closely related to the $qq$ potential from which they can be derived by
minimal substitution. Since the effect of the residual $qq$ potential is
clearly seen in the excited spectra of hadrons, one expects the corresponding
two-body currents to play an important role in various e.m. properties of
hadrons.

\begin{figure}[th]
\begin{center}
\epsfig{file=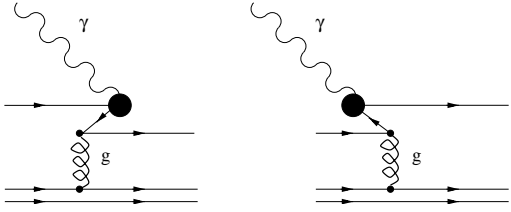,width=0.78\columnwidth,angle=0.,clip}
\caption{\label{OGE}
			Diagrams for OGE pair-currents. }
 \vspace{-1cm}
\end{center}
\end{figure} 
Both, the photon and the gluons interacting with quarks can produce $q\bar q$
pairs leading to pair-current contributions to e.m. quark current as provided
by OGE. The two-body terms we consider are depicted in Fig.~\ref{OGE}. The
nonrelativistic reduction of these diagrams leads to the following
configuration space e.m. current operators~\cite{Grabmayr1,Sanctis}:

\begin{eqnarray}
\label{rhooge}
\rho_{3q}^{(OGE)} = - i \frac{ \alpha_s}{16 m_q^3} \sum_{i < j}
{\bf \lambda}_i \cdot {\bf \lambda}_j \frac{ \mathcal{Q}_i}{r^3_{ij}}
\left[ e^{i {\bf q}\cdot {\bf r}_i} \Big({\bf q} \cdot
({\bf r}_{i} - {\bf r}_{j}) \right.
\hspace{0.cm} \nonumber \\
 +  \left.
\Big[{\bf \sigma}_i \times {\bf q}\Big] \Big[{\bf \sigma}_j \times
({\bf r}_{i}-{\bf r}_j)\Big] \Big)
+
(i \leftrightarrow j) \right] \hspace{0.8cm} \\
\label{joge}
{\bf j}_{3q}^{(OGE)} = - \frac{\alpha_s}{8 m_q^2} \sum_{i < j}
{\bf \lambda}_i \cdot {\bf \lambda}_j \frac{ \mathcal{Q}_i}{r^3_{ij}}
\hspace{3.3cm} \nonumber \\
\times ~
\Big[ e^{i {\bf q} \cdot {\bf r}_i}
\Big[({\bf \sigma}_i + {\bf \sigma}_j ) \times ({\bf r}_{i}-{\bf r}_j)\Big]
+
(i \leftrightarrow j) \Big] \hspace{0.2cm}
\end{eqnarray}

These OGE pair-currents describe a $q\bar q$ pair creation process induced by
the external photon with subsequent annihilation of the $q\bar q$ pair into a
gluon, which is then absorbed by an another quark. These currents are of
relativistic origin as reflected in the higher powers of $1/m_{q}$ as compared
to the one-body e.m. current operators.  Because the gluon does not carry any
isospin the OGE pair-current has the same isospin structure as the one-body
currents.  Eqs.(\ref{rhooge}) and~(\ref{joge}) result in the following
electric~$G^{(OGE)}_{E}$ and magnetic $G^{(OGE)}_{M}$ form factors:

\begin{eqnarray}
\label{OGE_E}
\left\{
\begin{array}{r}
G^{(OGE)}_{E_p} \\
G^{(OGE)}_{E_n}
\end{array}
\right\} &=& -\frac{\alpha_s}{m_q^3} ~ q ~ e^{-q^2 b^2 /24}
\left\{
\begin{array}{r}  1/3 \\  - 2/9   \end{array}
\right\} \mathcal{K}(q) ~~~~~~\\
\label{OGE_M}
\left\{
\begin{array}{r}
G^{(OGE)}_{M_p} \\
G^{(OGE)}_{M_n}
\end{array}
\right\}
&=&
\frac{\alpha_s}{m_q^2} ~ \frac{M_N}{q} ~ e^{-q^2 b^2 /24}
\left\{
\begin{array}{r}
2/3 \\
- 2/9
\end{array}
\right\} \mathcal{K}(q) ~~~~~~
\end{eqnarray}
The function $\mathcal{K}$ in the above expressions is:
\begin{equation}
\mathcal{K}(q) =
4 \pi \Big( \frac{1}{2 \pi b^2} \Big)^{3/2}
\int_{0}^{\infty} d r ~ e^{-r^2/(2 b^2)} j_1({q r}/{2})
\end{equation}
where $j_1({q r}/{2})$ is the spherical Bessel function.  The interaction of
the incoming photon with a $q\bar q$ pair  can be
considered as a point-like interaction or as being dominated by intermediate
vector mesons. The latter leads to an additional dipole form factor,
$ \ds F_{\gamma q \bar q}({\bf q}^2) = \Lambda_{\gamma q \bar q}^2/
\left(\Lambda_{\gamma q \bar q}^2 + {\bf q}^2\right)$, 
reflecting the extended structure of the $\gamma q \bar q$ vertex.
$\Lambda_{\gamma q \bar q}$ can be considered as a free parameter
or simply can be taken equal to the $\rho$-meson mass.  

{\it Results:}~~ In this work we consider the effect of the OGE pair-current
corrections to the NRCQM nucleon e.m. form factors, particularly for the
ratio~$\mu_{p} G^p_{E}/G^p_{M}$.  The ratio is calculated for a quark mass of
$m_q$=400~MeV and the respective quark core radius of $b$=0.5~fm.  In
Fig.~\ref{fig:gegmprot4} calculations with different $\alpha_s$ are shown to
indicate the sensitivity.  In the insert of Fig.~\ref{fig:gegmprot4} we show
results towards higher values of $Q^2$ for the best description of the present
data by $\alpha_s$ = 0.4 with (solid curve) and without Lorentz boost (dashed
curve).  Our results indicate that the ~$\mu_{p} G^p_{E}/G^p_{M}$ continues to
decrease and that it will cross zero at $Q^2\sim$8.1~GeV$^2$. From this, a
negative value of the ratio must be expected for the planed measurements in
JLAB at $Q^2\sim$9~GeV$^2$.  Deviations could be explained due to an extended
$\gamma q \bar q$ vertex as demonstrated by using a $\Lambda_{\gamma q \bar
q}$=770~MeV (dotted line). The introduction of such states does not affect
very much our results up to $\sim$10~GeV$^2$, but strongly influences the
behaviour of $\mu_{p} G^p_{E}/G^p_{M}$ for higher $Q^2$.  For quark masses in
the range $m_q\sim313\div400$~MeV and bag radii of $b\sim0.4\div0.6$~fm one
can find also a good description of the data with reasonable values for
$\alpha_s\sim0.2\div0.6$~\cite{Sanctis}.  However, these are not able to
reproduce the $N-\Delta$ mass splitting in the case of pure OGE. It seems
likely that the observed mass splitting is the result of a linear combination
of the pion-loop contributions and OGE~\cite{Thomas_book}. In this sense
pionic contributions could produce the desirable effect of reducing the size
of the strong coupling constant~$\alpha_s$, needed for the reproduction of the
~$\mu_{p} G^p_{E}/G^p_{M}$.
\begin{figure}[bt]
\begin{center}
\epsfig{file=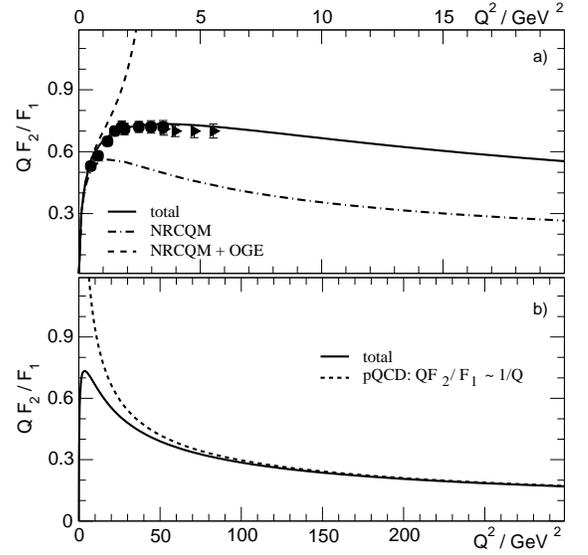,width=0.85\columnwidth,angle=0.,clip} 
\caption{\label{fig:qf2f1-highQ}
The $Q F_2/F_1$ ratio for the proton and its high $Q^2$ behavior. 
The quark and gluon contributions are shown~in~a).
}
\end{center}
\vspace{-0.7cm}
\end{figure}

The ratio $\mathcal{F}^p_2/\mathcal{F}^p_{1}$ can be directly derived from
$G^p_{E}/G^p_{M}$. It is predicted in Ref.~\cite{Miller} to be constant for
values of $Q^2$ up to 20~GeV$^2$ and it is understood as a result of the
Melosh transformation, which reflects relativistic effects. Our results are
shown in Fig.~\ref{fig:qf2f1-highQ}.  The ``kinematical'' background formed by
the naive CQM results (dot-dashed curve), $\mathcal{F}^p_2/\mathcal{F}^p_{1} =
1/(1+ \kappa_p Q^2/4M^2_M)$, underestimates the data and is not affected by
the Lorentz boost, a failure which is overcompensated when adding the OGE
pair-currents (dashed curve). It is due to the Lorentz boost (solid curve)
acting on the OGE currents to reproduce the flattening in
$Q\mathcal{F}^p_2/\mathcal{F}^p_{1}$. Following Ref.~\cite{MillerFrank}, we
also study the high $Q^2$-behavior. The ratio falls for asymptotic values of
$Q^2$ as $Q\mathcal{F}^p_2/\mathcal{F}^p_{1} \sim 1/Q$, and allows to make a
smooth transition to the scaling behavior as expected from
pQCD~\cite{Lepage}. In Ref.~\cite{MillerFrank} the ratio $Q
\mathcal{F}^p_2/\mathcal{F}^p_{1}$ falls less quickly as in our case and in
pQCD, both stated a notion of the hadron helicity conservation.  We also
confirm the statement of Ref.~\cite{MillerFrank}, that a plateau seen in
Fig.~\ref{fig:qf2f1-highQ} is the result of a broad maximum occuring near
$Q^2\sim10$~GeV$^2$.

\begin{figure}[t]
\begin{center}
\epsfig{file=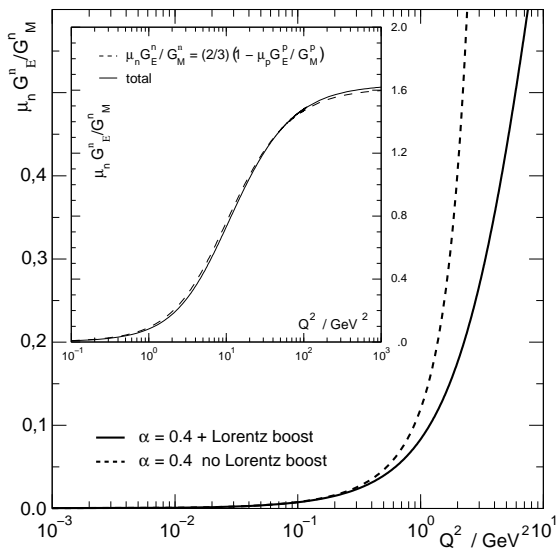,width=0.85\columnwidth,angle=0.,clip}
\caption{\label{fig:gegmneut}
Our predictions with and without Lorentz boost 
for the ratio $\mu_{n} G^n_{E}/G^n_{M}$ of the neutron. The insert gives the
ratio for a larger Q$^2$ range in comparison with Eq.(\ref{eq:approx}).}
\end{center}
\vspace{-0.7cm}
\end{figure}

Recent experimental progress in using polarized nuclear targets will allow to
obtain the neutron ratio $G_E^n/G_M^n$. As well known in the $SU(6)$ limit
$G_E^n(Q^2)$ is zero~\cite{IsgurNeutron}.  We can treat $G_E^n(Q^2)$ as a
result of the residual OGE-force in the form of gluonic currents and with the
assumption that the OGE pair-currents are the driving mechanism of the scaling
law violation we can calculate the neutron ratio $G_E^n/G_M^n$ using the best
results for the proton. The results are shown in
Fig.~\ref{fig:gegmneut}. Recombining
Eqs.~\ref{OB_E}, \ref{OB_M}, \ref{OGE_E} and \ref{OGE_M}
leads to a simple approximate result in analytic form between $G_E^n/G_M^n$
and that of the proton,~ $G_E^p/G_M^p$:
\begin{equation}\label{eq:approx}
\mu_n G_E^n/G_M^n \simeq \frac{2}{3} ~(1 - \mu_p G_E^p/G_M^p),
\end{equation}
which works remarkably well from low up to very high $Q^2$, and actually
insensitive to the choice of the parameters (insert Fig.~\ref{fig:gegmneut}).

In conclusion, we would like to mention that the internal dynamics of the
nucleon are much more complex than we have presented in this work.  First of
all it is interesting to examine the effect of nonvalence Fock
states~\cite{KG}:
\begin{equation}
\Psi_N =  \left(
\begin{array}{c}
\Psi(3q) \\
\Psi(3q + q \bar q)
\end{array}
\right)
\end{equation}
reflecting $q \bar q$ fluctuations of the constituent quarks. This question is
closely related to the 
possible role of the mesonic cloud.

\begin{acknowledgments}
Very useful discussions with V.I. Kukulin are gratefully acknowledged.  This
work was supported by the Deutsche Forschungsgemeinschaft under contracts
Gr1084/3, He2171/3 and GRK683.
\end{acknowledgments}


\end{document}